\documentclass[preprint,11pt]{elsarticle}

\usepackage[top=2.5cm,bottom=2.5cm,left=2.6cm,right=2.6cm,a4paper]{geometry}
\usepackage{graphicx}
\usepackage{dcolumn}
\usepackage{bm}
\usepackage{epstopdf}
\usepackage{mathrsfs}
\usepackage{amsmath,amssymb,amsfonts}
\usepackage{latexsym}
\usepackage{color}
\usepackage{slashed}
\usepackage{nicefrac}
\usepackage{hyperref}
\usepackage[utf8]{inputenc}
\usepackage[normalem]{ulem}

\def\bs{\boldsymbol} 

\def\bdel{\bs\partial}
\def\sst{\scriptscriptstyle}
\newcommand{\eqn}[1]{Eq.~\eqref{#1}}

\long\def\comment#1{ }

\newcommand{\onehalf}{{\nicefrac{1}{2}}}

\def\0{{\boldsymbol 0}}

\def\k{{\boldsymbol k}}

\def\n{{\boldsymbol n}}
\def\x{{\boldsymbol x}}
\def\y{{\boldsymbol y}}
\def\p{{\boldsymbol p}}
\def\r{{\boldsymbol r}}

\def\beps{{\boldsymbol \epsilon}}

\def\tmat{ \text{\bf t}}
\def\tmat{ \textbf{t}}

\def\ti{ t}
\def\tf{\bar t}
\def\tend{L}
\def\tform{{t_\text{f}}}
\def\tdecoh{t_\text{d}}
\def\tbroad{t_\text{broad}}

\def\tbr{t_\text{broad}}

\def\rmR{{\rm Re}}

\def\abar{{\rm \bar\alpha}}

\def\qhat{\hat q}

\def\thetad{\theta_\text{d}}
\def\thetabr{\theta_\text{broad}}
\def\thetaL{\theta_\text{\tiny L}}

\def\med{\text{med}}
\def\vac{\text{vac}}
\def\em{\text{em}}
\def\in{\text{in}}
\def\out{\text{out}}

\def\Gc{\mathcal{G}}

\def\Mc{\mathcal{M}}

\def\Dc{\mathcal{D}}

\newcommand{\beq}{\begin{eqnarray}}
\newcommand{\eeq}{\end{eqnarray}}
\newcommand{\be}{\begin{eqnarray*}}
\newcommand{\ee}{\end{eqnarray*}}
\newcommand{\bel}{\begin{align}}
\newcommand{\eel}{\end{align}}

\newcommand{\dd}{{\rm d}}
\newcommand{\rme}{{\rm e}}

\newcommand{\rmtr}{{\rm tr}}

\newcommand{\nn}{\nonumber\\ }

\newcommand{\gettitle}{Mapping collinear in-medium parton splittings}
\hypersetup{
	colorlinks,
	linkcolor={magenta},
	citecolor={blue},
	urlcolor={blue},
	pdftitle={\gettitle},
	pdfauthor={Dominguez, Milhano, Salgado, Tywoniuk, Vila},
	pdfkeywords={QCD jets} {Jet quenching},
	bookmarksopen=true,
	bookmarksopenlevel=2,
	bookmarksnumbered=true
}

\usepackage{color,soul}
\usepackage{xcolor}
\usepackage[normalem]{ulem}

\begin{document}
\begin{frontmatter}
\title{Mapping collinear in-medium parton splittings}

\author[1]{Fabio Dom\'inguez} 
\author[2,3]{Jos\'e Guilherme Milhano}
\author[1]{Carlos A. Salgado}
\author[4]{Konrad Tywoniuk}
\author[1]{V\'ictor Vila}

\address[1]{Instituto Galego de F\'isica de Altas Enerx\'ias IGFAE, Universidade de Santiago de Compostela, E-15782 Galicia-Spain}
\address[2]{LIP, Av. Prof. Gama Pinto, 2, P-1649-003 Lisbon, Portugal}
\address[3]{Instituto Superior T\'ecnico (IST), Universidade de Lisboa, Av. Rovisco Pais 1, 1049-001, Lisbon, Portugal}
\address[4]{Department of Physics and Technology, University of Bergen, 5020 Bergen, Norway}
\date{\today}

\begin{abstract}
We map the spectrum of $1\to 2$ parton splittings inside a medium characterized by a transport coefficient $\hat q$ onto the kinematical Lund plane, taking into account the finite formation time of the process. We discuss the distinct regimes arising in this map for in-medium splittings, pointing out the close correspondence to a semi-classical description in the limit of hard, collinear radiation with short formation times. Although we disregard any modifications of the original parton kinematics in course of the propagation through the medium, subtle modifications to the radiation pattern compared to the vacuum baseline can be traced back to the physics of color decoherence and accumulated interactions in the medium. We provide theoretical support to vacuum-like emissions inside the medium by delimiting  the  regions of phase space where it is dominant, identifying also the relevant time-scales involved. The observed modifications are shown to be quite general for any dipole created in the medium. 
\end{abstract}

\begin{keyword}
perturbative QCD \sep jet physics \sep jet quenching
\end{keyword}

\end{frontmatter}

\section{Introduction}

Jet quenching, the modification of jet observables in the presence of a QCD medium, is arguably the most versatile experimental tool to characterize the hot and dense system created in heavy-ion collisions, see e.g. \cite{dEnterria:2009xfs,Majumder:2010qh,Mehtar-Tani:2013pia}. In the last  20 years experiments at RHIC \cite{Adams:2005dq,Adcox:2004mh} and then the LHC \cite{Aamodt:2010jd,Khachatryan:2016odn,Chatrchyan:2011sx,Aad:2010bu,Abelev:2013kqa} found a strong suppression of particles produced at high transverse momentum,  the most direct predictions of  energy loss, one of the clearest signatures of the presence of jet quenching dynamics. The large kinematical reach of the LHC, and the much larger integrated luminosities expected for the near future, allow to adapt and design completely new jet tools \cite{Acharya:2017goa,Sirunyan:2018gct}, originally developed for the proton-proton program, with a much more differential access to different properties of the medium. Two examples are the access to different energy scales in the medium properties, notably, the access to short distances to measure the properties of its quasiparticle content \cite{DEramo:2012uzl} and the access to the space-time evolution of the whole system \cite{Apolinario:2017sob}, including its initial stages \cite{Andres:2019eus}. One essential ingredient for a correct interpretation of the data is a good control of the splitting process in the relevant energy- or time-scale under investigation. Interestingly, the new jet tools mentioned above can be used to isolate, or at least make it cleaner, these different scales \cite{Andrews:2018jcm}. 

The problem of elementary parton splittings is important in many aspects of high-energy physics. Most prominently, it allows to resum final-state emissions that accompany hard processes at colliders (a similar framework exists for resummation of initial-state radiation, but here we focus on the former). This is manifested experimentally as sprays of collimated particles, i.e. a jet. The fundamental splitting vertices together with a calculation of the available phase space are the ingredients that enter the formulation of a Monte Carlo parton shower.

For processes involving soft gluon radiation, one often invokes a strong separation of scales that allows to define a classical current. 
In a diagrammatic language, the current represents high-energy particles that act as sources of soft gluons and originate from a espacial position that is fixed in both the amplitude and its complex conjugate. This method has been shown to provide an economical description of both initial- and final-state emissions in the presence of a nuclear medium, e.g. see \cite{MehtarTani:2006xq}.
Similarly, the interference pattern off multiple emitters was studied assuming an instantaneous splitting of the current, giving rise to the so-called antenna radiation pattern \cite{MehtarTani:2010ma,MehtarTani:2011tz,MehtarTani:2012cy,CasalderreySolana:2011rz}. This picture was further corroborated within a diagrammatic calculation of the two-gluon emission spectrum in the limit of strong ordering of their respective emission times \cite{Casalderrey-Solana:2015bww}. Recently, both Monte-Carlo studies \cite{Milhano:2015mng,Casalderrey-Solana:2016jvj} and analytical calculations \cite{Mehtar-Tani:2017web,Caucal:2018dla}, have highlighted the role of jet fluctuations that arise from in-medium splittings on observables that are sensitive to energy loss in heavy-ion collisions.

In this work we compute, within a diagrammatic approach, a real and collinear parton splitting inside a color deconfined medium and study the set of medium-induced modifications that arise from allowing this splitting to occur at a finite distance within the medium. Our discussion is most clearly cast in the context of a final-state color-singlet splitting, i.e. $\gamma \to q \bar q$, but remains valid for generic splitting processes involving a total color charge. We systematically implement the high-energy limit in our calculations, that reduces the complexity of the problem to a semi-classical picture of partons propagating along well-defined trajectories. Although we disregard any modifications of the original parton kinematics in course of the propagation through the medium, subtle modifications to the radiation pattern compared to the vacuum baseline can be traced back to the impact of physical time-scales in the medium, related to color decoherence and broadening.\footnote{Technically, these two processes relate to the physics of two types of dipole survival probabilities: the former, to a dipole existing entirely in the amplitude (or complex conjugate amplitude), and the latter, to a fictitious dipole consisting of one trajectory in the amplitude and another in the complex conjugate amplitude.}

The time-like separation of the splitting vertices in, respectively, the amplitude and the complex-conjugate amplitude gives rise to two- (dipole) and four-point (quadrupole) correlators of Wilson lines that resum medium interactions. These correspond to the survival probabilities of the two- and four-parton configurations at various stages of the process under consideration. It is crucial to note that in the absence of this separation these correlators collapse to unity, leaving no imprint on the splitting process. The appearance of the quadrupole, describing the propagation of the pair from formation time to the end of the medium, is essential since it accounts for the accumulated effects of medium interactions over long distances.

Let us also clarify what we mean by the decoherence of the dipole. In earlier works, where the dipole was assumed to form quasi-instantaneously close to the origin, color decoherence was shown to introduce a new timescale that governs the spectrum of subsequent soft emissions \cite{MehtarTani:2011tz,CasalderreySolana:2011rz,MehtarTani:2012cy}. This comes about because the interference pattern between the radiation off each of the dipole constituents depends on the color coherence of the pair. In the current setup, we study in detail the \textsl{formation of the dipole} itself, for the moment without considering further radiative processes and ask the simple question of when and how such dipoles are formed. An important property to understand in this context is the locality of the splitting, i.e. whether the properties of the parton pair are determined at the moment of formation or whether those still can undergo modifications over large distances in the medium. Our results point to the importance of both regimes and quantify them in terms of logarithmic phase space. In close analogy to studies of dipole radiation patterns in vacuum, it is very helpful to map the kinematics of the formed dipoles onto the kinematical Lund plane \cite{Andersson:1988gp}. Filling the Lund plane using jet de-clustering techniques in proton-proton and heavy-ion collisions has recently attracted a lot of attention \cite{Andrews:2018jcm,Chien:2018dfn,Dreyer:2018nbf,Cunqueiro:2018jbh,Marzani:2019hun}.

Our final results for the emission spectrum in the presence of a medium can be cast in the form,
\beq
{\frac{\dd N^\med}{\dd z \,\dd \theta}}\bigg/{\frac{\dd N^\vac}{\dd z\, \dd \theta}} = 1+ F_\med \,,
\eeq
where we have explicitly factorized the medium-induced cross-section into the vacuum cross-section and the medium-induced modification. The function $F_\med$ encodes all information relative to  the medium modification factor associated with the parton $1\to 2$ splitting function. We discuss the relevant approximations in the high-energy limit that allow to simplify the description in \autoref{sec:semi-classical} and derive this expression in \autoref{sec:spectrum}. In practice, this factorization allows for the straightforward discrimination of medium effects, as will be fully explained in \autoref{sec:spectrum}. We discuss the relevant time-scales contained in the spectrum in \autoref{sec:scales} and draw the phase space for the process under consideration in \autoref{sec:LundDiag}. Then, in \autoref{sec:numerics}, we discuss numerical results that largely verify the preceding analysis. The steps needed to generalize the process under consideration to be valid for arbitrary splitting processes, involving e.g. color-charged dipoles etc., are outlined in \autoref{sec:beyondsinglet} and, finally, we present an outlook in \autoref{sec:conclusions}.

\section{Implementing the semi-classical limit}
\label{sec:semi-classical}

\begin{figure}[t!]
\centering
\includegraphics[width=0.55\textwidth]{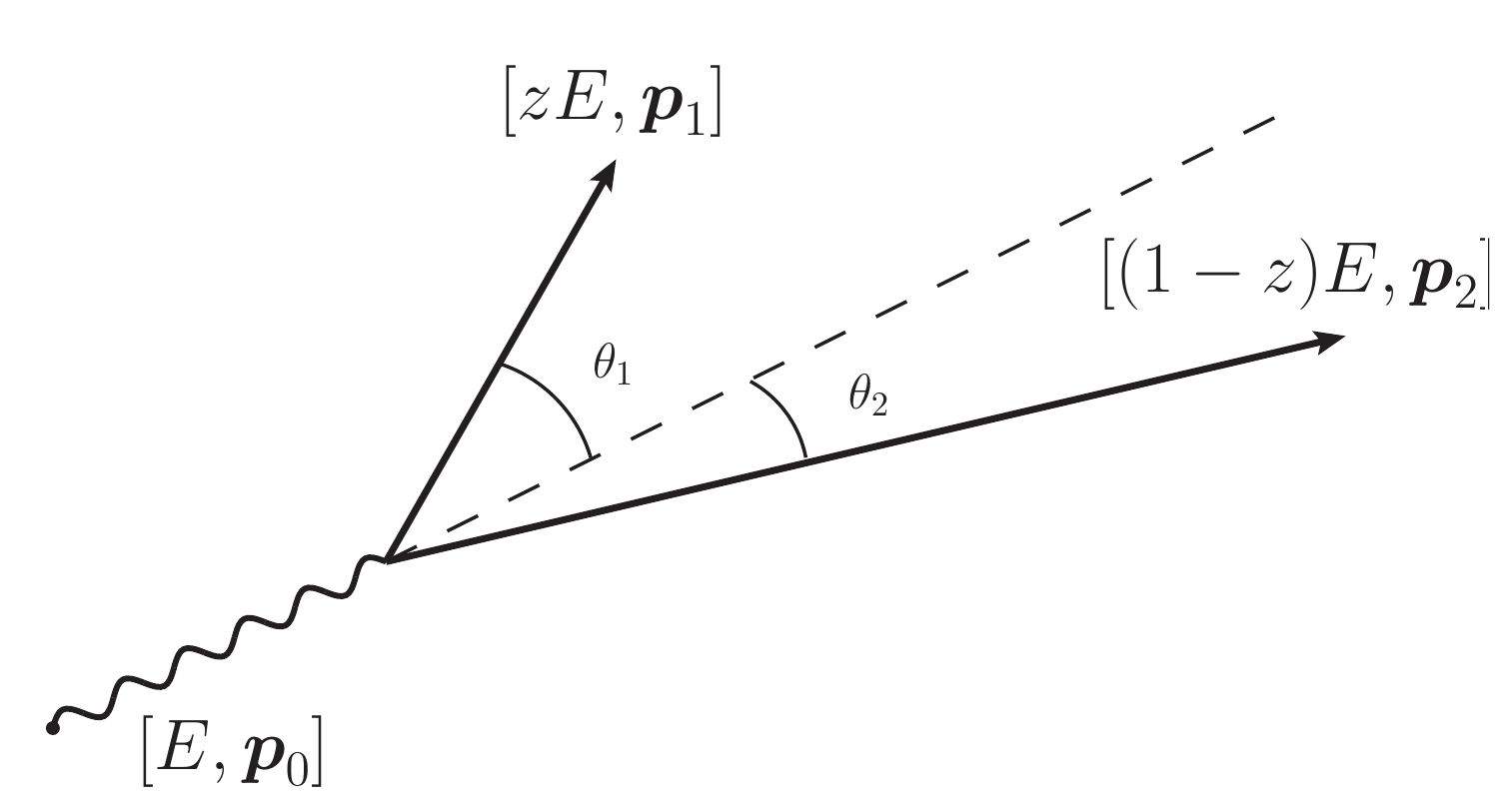}
\caption{Kinematics of the parton splitting process. The dipole opening angle is $\n_{12} = \n_1- \n_2$, with $|\n_1| = \theta_1$, $|\n_2| = \theta_2$ and $|\n_{12}| = \theta$.}
\label{fig:kinematics}
\end{figure}

Let us consider the splitting of a ``parent'' parton, with momentum $\vec p_0 = [E,\p_0]$, into two ``daughter'' partons, with final-state kinematics
\beq
\vec p_1 = [zE,\p_{1}] \,, \;\; \vec p_2 = [(1-z)E,\p_{2}] \,,
\eeq
where $\p_0 = \p_1 + \p_2$ from conservation of transverse momentum.\footnote{Our notation refers explicitly to light-cone (LC) kinematics, with $x^\pm = (x^0 \pm x^3)/\sqrt{2}$. Hence, $t\equiv x^+$ stands for LC time and $E\equiv p^+$ is LC energy. Throughout, the minus ($-$) component of the momenta has been integrated out, giving rise to an explicit time dependence.} For the moment, we will focus on the production of a color singlet final-state, in particular the splitting of a photon into a quark-antiquark pair, $\gamma \to q\bar q$, where the photon is transversely polarized. This is illustrated in \autoref{fig:kinematics}. The partial amplitude $\Mc$ for such a process, stripped of the Born-level production amplitude, reads (see also \cite{Wiedemann:2000ez})
\begin{align}
\label{eq:AmplitudeMom_1}
\Mc_{\gamma \rightarrow q \bar{q}} &= \frac{e}{E} \rme^{i \frac{\p_1^2}{2zE} \tend + i \frac{\p_2^2}{2(1-z)E}\tend } \int_0^\infty \dd \ti \ \int_{\k_1,\k_2} \left[\mathcal{G}(\p_{1},\tend ; \k_{1}, \ti | zE) \, \mathcal{\bar{G}}(\p_{2}, \tend; \k_{2}, \ti | (1-z)E) \right]_{ij} \nn
&\times \gamma_{\lambda,s,s'}(z) \k \cdot \beps_\lambda^\ast \, \mathcal{G}_{0}(\k_1 + \k_2,\ti | E)
\end{align}
where $e$ is the QED coupling constant and $\gamma_{\lambda,s,s'}(z) =  i \delta_{s,-s'} [z \delta_{\lambda,s} - (1-z) \delta_{\lambda,-s}]/\sqrt{z(1-z)}$ is the quark-gluon splitting vertex with $\lambda = \pm 1$ and $\k \equiv (1-z) \k_1 - z \k_2$. Throughout, we implement the notations $\int_{\x} \equiv \int \dd^2 \x$ for transverse coordinate and $\int_{\k} \equiv \int \dd^2 \k/(2\pi)^2$ for transverse momentum integrations. In this expression, $\mathcal{G}$ and $\mathcal{\bar G}$ represent the dressed retarded propagators for the quark and the antiquark, that incorporate an adiabatic turn-off at large times, see e.g. \cite{Wiedemann:2000za} for details on this regulator. As usual for high-energy processes, we have assumed that interactions with the medium only exchange transverse momentum. Hence the momenta $\k_1$ and $\k_2$ correspond to the transverse momentum sharing immediately after splitting.

The fully dressed propagator in the momentum space transforms to configuration space according to,
\begin{align}
\label{eq:G-FourierTransform}
\mathcal{G}( \p_{1}, t_{1}; \p_{0},t_{0}) & = \int_{\x_1,\x_2}  \rme^{ -i \p_{1} \cdot \x_{1}+ i \p_{0} \cdot \x_{0}} \mathcal{G}(\vec x_{1}, \vec x_{0} ) \,,
\end{align}
where we have suppressed the color and energy indices. In configuration space, $\Gc(\vec x_1, \vec x_0)$ is described by a 2+1 dimensional path integral along the trajectory of the particle, 
\beq
\label{eq:dressed-propagator-configuration}
\mathcal{G} (\vec x_1, \vec x_0) = \int_{\r(t_0) = \x_0}^{\r(t_1) = \x_1} \Dc \r \, \exp \left[i\frac{E}{2} \int_{t_0}^{t_1} \dd s \, \dot \r^2 \right] V(t_1,t_0 ; [\r]) \,,
\eeq
where the (conserved) energy $E$ acts as a ``mass''. Here, $V(t_1,t_0;[\r])$ is a Wilson line in the fundamental representation at (possibly fluctuating transverse)  position $\r(t)$, 
\beq
V(t_1,t_0;[\r]) = \mathcal{P} \exp \left[ig \int_{t_0}^{t_1} \dd t \, \tmat^a A^{-,a}(t,\r(t))\right] ,
\eeq
where $\tmat^a$ is a color matrix in the fundamental representation and $A^{-,a}(t,\r)$ is a background field describing interactions with the medium.\footnote{The background field is boosted in the opposite direction to the projectile and, hence, it is contracted at $x^- = 0$. This, in turn, guarantees conservation of longitudinal momentum in the propagator and permits the representation in \eqn{eq:dressed-propagator-configuration}.} The antiquark propagator $\mathcal{\bar G}(\vec x_1; \vec x_0)$ is defined analogously to \eqref{eq:dressed-propagator-configuration} with the substitution $V(t_1,t_0;[\r]) \to V^\dagger (t_1,t_0;[\r])$. For propagation outside of the medium, these propagators reduce to 
\beq
\mathcal{G} (\p_1, \tend; \k_1, \ti | E) \big|_{\ti>\tend} = (2\pi)^2 \delta(\k_1 - \p_1) \,\mathcal{G}_{0}(\p_1,\tend - \ti|E) \,,
\eeq
and analogously for $\mathcal{\bar G}$, where $\mathcal{G}_0(\k,t|E) = \rme^{-i\frac{\k^2}{2E} t}$. This corresponds also to the photon propagator in \eqn{eq:AmplitudeMom_1}.
 
The goal of this work is to focus on the limit of hard splittings in the medium, i.e. splittings with formation times much shorter than the typical time-scales of the medium, where we expect a ``semi-classical'' picture to dominate the cross section. We make this statement more precise and map out the relevant region in phase space in \autoref{sec:scales} and \autoref{sec:LundDiag}. This is in contrast to the limit of medium-induced branching \cite{Baier:1996kr,Zakharov:1997uu,Wiedemann:2000za,Gyulassy:2000er,Wang:2001ifa,Arnold:2002ja,Blaizot:2012fh,Apolinario:2014csa,Sievert:2019cwq}, where one investigates emissions with transverse momenta dominated by interactions with the medium, i.e. $p_\perp \sim \sqrt{\hat q z E}$. We will work in the high-energy limit, i.e. where formally the energy of the particles is infinite, $E \to \infty$, but we will keep track of the finite momentum sharing fraction $z$. It turns out that we need to consider two separate steps in order to establish this correspondence, which we proceed to outline below. The first step fixes the trajectories of the particles to follow classical trajectories that are determined by the kinematics of the process while the second one fixes a common reference point for the pair in transverse coordinates.

By hard emissions, we explicitly mean that both partons have energy large enough so that the change in transverse position due to scattering with the medium can be neglected and the propagation follows basically straight lines. This contribution can be isolated by considering the so-called eikonal expansion of the propagator \eqref{eq:dressed-propagator-configuration}  \cite{Altinoluk:2014oxa,Altinoluk:2015gia}. Its zeroth-order term, which neglects further transverse momentum broadening in the medium, turns out to be
\beq
\label{eq:Propagator_Eikonal}
\Gc^{(0)} (\vec x_1, \vec x_0)  = \Gc_0(\x_1 - \x_0,t_1-t_0) V(t_1,t_0;[\x_{\rm cl}]) \,,
\eeq
where $\x_{\rm cl}(t) = \frac{t_1-t}{t_1-t_0}\x_0 + \frac{t-t_0}{t_1-t_0} \x_1$ is the classical path. Taking the Fourier transform, see \eqn{eq:G-FourierTransform}, and after some manipulations, we find that the propagator in mixed representation becomes
\begin{align}
\label{eq:prop-1}
\Gc^{(0)}(\p_1,t_1;\p_0,t_0) &= \rme^{-i \frac{\p_1^2}{2E}(t_1-t_0)} \int_{\y_0,\y_1} \, \rme^{-i (\p_1 - \p_0) \cdot \y_0} \,\frac{E (t_1-t_0)}{2\pi i} \rme^{i \frac{E(t_1-t_0)}{2} (\y_1 - \n )^2} \nn
& \times V(t_1,t_0;[\y_0 + (t-t_0)\y_1]) \,,
\end{align}
where $\n \equiv \p_1/E$. In the ``semi-classical'' limit $E(t_1-t_0) \gg 1$ (corresponding to the formal limit $\hbar \to 0$) the heat-kernel in \eqref{eq:prop-1} reduces to a delta function of its argument, $\lim_{\varepsilon \to 0 } \rme^{- \x^2/\varepsilon} /(\pi \varepsilon) = \delta(\x)$. In particular, we demand that $E \gg L^{-1}$. This step converges the particles path onto the classical trajectory, and the propagator becomes
\beq
\label{eq:propagator-nobroad}
\Gc^{(0)}(\p_1,t_1;\p_0, t_0) = \rme^{-i \frac{\p_1^2}{2E}(t_1-t_0)} \int_\x \, \rme^{-i(\p_1 - \p_0)\cdot \x} V(t_1,t_0;[\x + (t-t_0)\n]) \,,
 \eeq
see also \cite{Mehtar-Tani:2017ypq}. This expression corresponds to the S-matrix of an energetic particle that traverses the medium, see e.g. \cite{CasalderreySolana:2007zz}.

Hence, in this first step we have removed all effects of {\it non-eikonal} broadening in the medium, i.e. that associated to the fluctuations in the transverse position, after the pair has been created. However, the initial position of the trajectory of each leg is not fixed by \eqref{eq:propagator-nobroad}. This leads to a ``smearing'' of the antenna initial position in transverse space. We will briefly return to this detail in~\ref{sec:beyondclassical}. For physical processes happening at large times from the initial position we can treat the initial position of the Wilson line as a small correction. Then we find
\beq
\label{eq:PropagatorSemiClassical}
\Gc^{(0)}(\p_1,t_1;\p_0, t_0) =(2 \pi)^{2} \delta(\p_{0}-\p_{1}) \, \rme^{-i\frac{\p_{1}^{2}}{2E}(t_1-t_0)} V( t_{1}, t_{0}; \big[ \n t\big])
\eeq
for the quark propagator. This ensures that the Wilson lines accompanying the two hard particles always are initiated at the same initial transverse position and time.

Replacing the propagators in \eqn{eq:AmplitudeMom_1} with \eqref{eq:PropagatorSemiClassical} and considering the splitting {\it inside} the medium, $0 < \ti < \tend$, the amplitude becomes
\beq
\label{eq:in-amplitude-1}
\Mc^{\rm in}_{\gamma \rightarrow q \bar{q}} =\frac{e}{E}\gamma_{\lambda,s,s'}(z) \, \p\cdot \beps_\lambda^\ast \int_0^\tend \dd \ti \,\exp\left(-i \frac{\tend- \ti}{\tform} \right) \Big[V_{1}(\tend, \ti) V^{\dagger}_{2}(\tend,\ti)\Big]_{ij},
\eeq
where now $\p\equiv (1-z) \p_1 - z \p_2$ is only related to the final-state momenta. In this expression, we identify the quantum-mechanical formation time
\beq
\label{eq:FormationTime}
 \tform =\frac{2z(1-z)E}{\p^{2}} \,.
 \eeq
This second step completes the {\it semi-classical} approximation where the particles are propagating along trajectories determined by their kinematics. The two Wilson lines are associated with each of the dipole constituents, such that e.g. $V_i(\tf,\ti)\equiv V(\tf,\ti;[\r_i(s)])$ with $\r_i(s) = \n_i s$. Explicitly, $\n_1 = \p_1/[zE]$ and $\n_2 = \p_2/[(1-z)E]$. For emissions taking place {\it outside} of the medium, $t> L$, on the other hand, we can explicitly perform the integration over the splitting time. The amplitude reads then
\beq
\label{eq:OutAmplitude}
\Mc^{\rm out}_{\gamma \to q\bar q} = \delta_{ij} \,\frac{i \, z(1-z)\,e}{E} \gamma_{\lambda,s,s'}(z) \frac{\p\cdot \epsilon^\ast_\lambda}{\p^2},
\eeq
where we have assumed the adiabatic turn-off prescription at large times mentioned above (consistent with the usual Feynman prescription in the vacuum propagators in momentum space). The amplitudes \eqref{eq:in-amplitude-1} and \eqref{eq:OutAmplitude} are written up to pure phase factors that cancel out in the cross sections. The full amplitude is simply the sum $\Mc_{\gamma \to q\bar q} = \Mc^{\rm in}_{\gamma \to q\bar q} + \Mc^{\rm out}_{\gamma \to q\bar q}$.

\section{Derivation of the spectrum}
\label{sec:spectrum}

\begin{figure}[t!]
\centering
\includegraphics[width=0.55\textwidth]{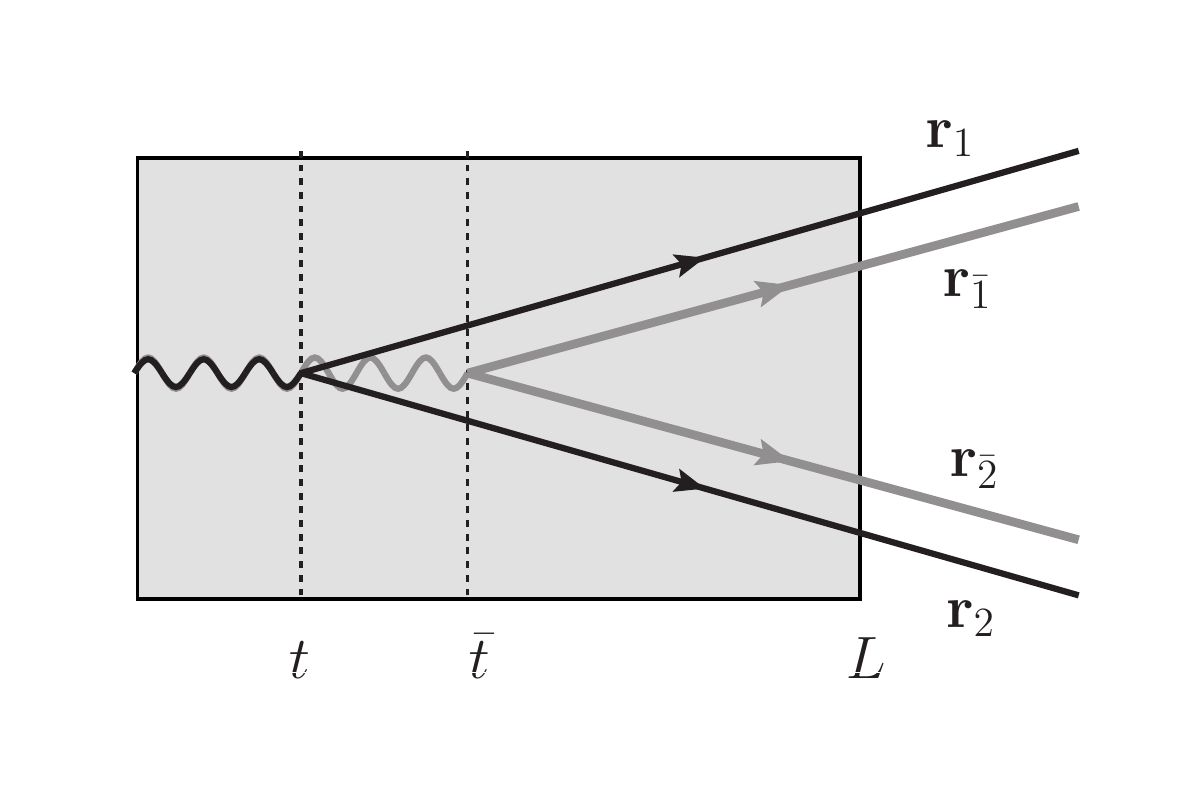}
\caption{The ``in-in'' contribution to the spectrum. Amplitude (black lines) and complex conjugate of the amplitude (grey lines) are plotted on top of each other to clearly show the different contributions: dipole (\ref{eq:specdipole}) in the region $(t,\bar t)$ and quadrupole (\ref{eq:specquadrupole}) in $(\bar t,L)$.}
\label{fig:photondipole}
\end{figure}

Moving to the computation of the spectrum of splittings, the following definitions will become very helpful. We note that in our setup the trajectories of the particles in the amplitude, denoted by $\r_1$ and $\r_2$, and complex-conjugate amplitude, denoted $\r_{\bar 1}$ and $\r_{\bar 2}$, are shifted only due the difference in the splitting time which vary independently, see \autoref{fig:photondipole}. The two-point function,
\beq
S_{\sst IJ}(t_1,t_0) \equiv \frac{1}{N_c} \langle \rmtr \,V_{\sst I} V^\dagger_{\sst J} \rangle \,,
\label{eq:specdipole}
\eeq
corresponds to the dipole cross section,\footnote{Throughout we adapt a notation where summation over common indices of adjacent color matrices is automatically summed over, and $(\tmat^a)^{ij} (\tmat^b)^{jk} \equiv [\tmat^a \tmat^b]^{ik}$, etc.} where  the extent of the Wilson lines is implicit from the left-hand side of the equation and $\{I,J \} = \{1,2,\bar 1, \bar 2\}$. Finally, the four-point function,
\beq
Q(t_1,t_0) = \frac{1}{N_c} \langle \rmtr \, V_1 V_2^\dagger V_{\bar 2} V_{\bar 1}^\dagger \rangle \,,
\label{eq:specquadrupole}
\eeq
is also referred to as a quadrupole. In the harmonic oscillator approximation and in a static medium of size $L$, for details see \cite{CasalderreySolana:2007zz},
\beq
S_{\sst IJ}(t_1,t_0) = \exp \left[ - \frac{1}{4} \qhat \int_{t_0}^{t_1} \dd s \, \r^2_{\sst IJ}(s) \right] \,,
\label{eq:hoa}
\eeq
where $\qhat \equiv C_F \hat{\overline q}$ is the transport coefficient for fundamental degrees of freedom and $\r_{\sst IJ} \equiv \r_{\sst I} - \r_{\sst J}$ describes the separation of the Wilson lines in transverse coordinate.
In the large-$N_c$ limit, the quadrupole can be written as \cite{Blaizot:2012fh,Apolinario:2014csa,Sievert:2019cwq},
\beq
\label{eq:quadrupole}
Q(t_1,t_0) = S_{1 \bar1}(t_1,t_0) S_{2\bar 2}(t_1,t_0) + \int_{t_0}^{t_1} \dd s\, S_{1 \bar1}(t_1,s) S_{2\bar 2}(t_1,s) \, T(s)\, S_{1 2}(s,t_0) S_{\bar 1\bar 2}(s,t_0) \,,
\eeq
where $T(s) = - \qhat[ \r_{12}^2(s) + \r_{\bar 1 \bar 2}^2(s) - \r_{1 \bar 2}^2 (s) - \r_{\bar 1 2}^2 (s)]/2 = - \hat q \r_{1\bar 1} (s) \cdot \r_{2 \bar 2} (s)$ is the transition amplitude. Above, the two terms in \eqref{eq:quadrupole} are the so-called factorizable and non-factorizable pieces of the quadrupole, and describe the propagation of two possible color configurations in the large-$N_c$ limit, under the constraint of conserving color at any given time. Concretely, the first term describes the propagation of two dipoles that correlate separately particle 1 and particle 2 in amplitude and complex conjugate (c.c.) amplitude so that they evolve independently. This is why this piece is called factorizable. The second, non-factorizable term involves a one-gluon exchange---described by the transition amplitude---that alters the color correlation of the system from an initial correlation of particles 1 and 2 in the amplitude and similarly in the c.c. at times $t< s$, to the uncorrelated case at times $t>s$.

Note that, for the problem at hand, the separations are either constant or grow linearly with time. In the high-energy limit, it is the separation of the splitting time in the amplitude $\ti$ and the splitting time in the c.c. amplitude $\tf$ that govern the exact trajectories of the Wilson lines. For reference, we list the relevant differences here,
\begin{align}
\label{distances}
\r_{1\bar 1}(s) &= \n_1 (\tf - \ti)\,, \; \; \r_{2\bar 2}(s) = \n_2 (\tf- \ti) \,, \\
\r_{12}(s) &= \n_{12}(s-\ti) \,, \; \; \r_{\bar 1 \bar 2}(s) = \n_{12}(s-\tf) \,,
\end{align}
where $\n_{12} \equiv \n_1 - \n_2$. Note also that $\n_{12} = \p /(z(1-z)E)$, where $\p = (1-z) \p_1 - z \p_2$ is the relative transverse momentum of the pair. Assuming a vanishing initial momentum $\p_0$, we can also deduce that $\n_1 = (1-z) \n_{12}$ and $\n_2 = -z \n_{12}$, with $\theta \equiv |\n_{12}|$ and $\theta_{1(2)} \equiv |\n_{1(2)}|$, see \autoref{fig:kinematics}. 
In this case, the transition amplitude takes the simple form
\beq
T(s) = -\qhat \, \n_1\cdot \n_2(\tf - \ti)^2 = - \hat q z(1-z) \n_{12}^2 (\tf - \ti)^2.
\eeq
 For future reference, we take note that  $\sum_{\lambda,s,s'} |\gamma_{\lambda,s,s'}(z)|^2 = [z^2+ (1-z)^2] /[z(1-z)] $.

The inclusive spectrum for the splitting process we are considering can be written as,
\beq
\frac{\dd N^{\rm med}}{ \dd z \dd \p^2} = \frac{1}{4 (2\pi)^2 \, z(1-z)} \langle \left| \Mc_{\gamma \to q\bar q} \right|^2 \rangle =\frac{1}{4 (2\pi)^2 \, z(1-z)} \langle \left| \Mc_{\gamma \to q\bar q}^{\rm in}+\Mc_{\gamma \to q\bar q}^{\rm out} \right|^2 \rangle \,,
\eeq
where the averaging of the amplitude also takes into account averaging over the ensemble of medium configurations. The total spectrum in the presence of a medium can be decomposed into three parts, $N^{\rm med} = N^{\rm in-in} + N^{\rm in-out} + N^{\rm vac}$. Here, the first contribution corresponds to a splitting taking place inside the medium in both the amplitude and the c.c. amplitude, the second contribution is an interference between an emission taking place inside the medium in the amplitude and outside the medium in the c.c. amplitude, or vice versa, and the last term corresponds to an emission outside the medium. We define the vacuum cross-section  from $ \langle |\Mc^\text{out}|^2\rangle $. It reads,
\beq
\label{eq:VacuumSpectrum}
\frac{\dd N^\vac}{ \dd z \,\dd \theta} = \frac{\alpha_\em}{\pi} \frac{P_{q\gamma}(z)}{\theta} \,,
\eeq
where we used that $\p^2 = [z(1-z)E\theta]^2$, $P_{q\gamma}(z) = n_f N_c [z^2+ (1-z)^2]$ being the relevant Altarelli-Parisi splitting function and $n_f$ is the number of quark flavors. Then, after simplifying, we can write for the ``in-in'' spectrum
 \begin{align}
\label{eq:inin-semiclassic}
\frac{\dd N^\text{in-in}}{\dd z \, \dd \theta} &= \frac{\dd N^\text{vac}}{\dd z \, \dd \theta}\; 2 \rmR \int_0^\tend \frac{\dd \ti}{\tform} \int_{\ti}^\tend \frac{\dd \tf}{\tform} \, \rme^{-i \frac{\tf - \ti}{\tform}} \,Q (\tend, \tf) S_{12}(\tf,\ti)  \,,
\end{align} 
where the quadrupole $Q(\tend,\tf) \equiv Q(\tend,\tf ; \ti)$ explicitly depends on the splitting time in the amplitude through the finite longitudinal shift of the long-distance propagators. The in-out spectrum reads,
\beq
\label{eq:inout-semiclassic}
\frac{\dd N^\text{in-out}}{\dd z \, \dd \theta} = - \frac{\dd N^\text{vac}}{\dd z \, \dd \theta}\; 2 \text{Im} \int_0^\tend \frac{\dd \ti}{\tform} \, \rme^{-i \frac{\tend-\ti}{\tform}} \, S_{12}(L,\ti) \,.
\eeq
Summing up all three contributions, the final spectrum takes the form
\beq
\label{eq:MediumSpectrum}
\frac{\dd N^\med}{ \dd z\, \dd \theta}  = \frac{\dd N^\vac}{ \dd z\, \dd \theta} \, \big(1+F_\med(z,\theta) \big)\,,
\eeq
where the medium modifications are encoded in the factor $F_\med$ that reads,
\beq
\label{eq:MediumModification}
F_\med = 2 \int^L_0 \frac{\dd \ti}{\tform} \left[ \int^L_{\ti} \frac{\dd \tf}{\tform} \, \cos\left(\frac{\tf-\ti}{\tform} \right) S_{12}(\tf,\ti) Q(\tend,\tf) - \sin \left( \frac{L - \ti}{\tform} \right) S_{12}(\tend,\ti) \right] \,,
\eeq
with $S_{12}(\tf,\ti) \equiv S_{12}(\tau)$ and $Q(\tend,\tf) \equiv Q(\tau_L,\tau)$ that only depend on the differences $\tau = \tf - \ti$ and $\tau_L = \tend-\tf$. Explicitly, these functions read
\begin{align}
\label{eq:FinalDipole}
S_{12}(\tau) &= \rme^{-\frac{1}{12}\hat q \theta^2 \tau^3} \,,\\
\label{eq:FinalQuadrupole}
Q(\tau_L,\tau)  &= \rme^{-\frac{1}{4} \qhat \xi \theta^2 \tau_L \tau^2} + T(\tau) \int^{\tend}_{\tf} \dd s\, \rme^{-\frac{1}{4} \qhat \xi \theta^2 (L - s) \tau^2} \rme^{-\frac{1}{12}\hat q \theta^2 [(s-\tf)^3+(s-\ti)^3 - \tau^3]} \,,
\end{align}
where $s \geq \tf \geq \ti$ and we defined $\xi = (1-z)^2+z^2$. The factorization property in \eqn {eq:MediumSpectrum} stems from the fact that the kinematics of the dipole is not modified after it has been created. In contrast, for medium-induced branching the daughter particles experience additional  momentum broadening from to non-eikonal contributions both during their formation time and afterwards \cite{Blaizot:2012fh,Apolinario:2014csa,Sievert:2019cwq}.
The dependence on the initial energy does not factorize completely on the right-hand-side of the equation due to the explicit dependence on the formation time $\tform$.

The process described by \eqn{eq:MediumModification} contains two stages. To be accurate, for the in-out term, see \eqref{eq:inout-semiclassic}, only the first stage plays a role. The first stage is governed by the dipole cross section $S_{12}(t_1,t_0)$ that appears in both terms in \eqn{eq:MediumModification}. It can be interpreted as a survival probability of a (virtual) dipole consisting of the daughter particles 1 and 2, with a dynamical transverse size $b_\perp(t) \sim \theta t$, that exist during the time interval $\Delta t = t_1-t_0$. We will refer to this stage of the process as the ``decoherence'' of the pair. The quadrupole $Q(\tau_L,\tau)$ plays only a role for the ``in-in'' term, see \eqn{eq:inin-semiclassic}. Looking in detail, the first term in \eqref{eq:FinalQuadrupole} describes the survival probability of a (real) dipole with fixed transverse size $b_\perp \sim \theta \tau$ at the moment of formation, propagating the remaining distance to the end of the medium. 
Therefore, we will refer to this part of the dynamics as the broadening of the dipole. As mentioned before, in our approximation this broadening does not receive contributions from the non-eikonal fluctuations that change the transverse position of the propagating particles off their classical paths. The non-factorizable piece, given by the second term in \eqn{eq:FinalQuadrupole}, is typically a small correction. For example, the transition amplitude $T(\tau) \sim z \theta^2 \tau^2$ becomes approximately $T(\tau) \sim \tform/E$ for short-formation times, i.e. when $\tau \sim \tform \ll L$ (see discussion in \autoref{sec:scales}). It also vanishes in the soft limit $z\to 0$ and $\tau \sim$ const.

The terms in \eqref{eq:MediumModification} correspond, respectively, to the cases when the splitting occurs inside the medium in both the amplitude and its complex conjugate, $\propto \langle | \Mc^\in_{\gamma \to q \bar q}|^2\rangle$ (first term), already referred to as an ``in-in'' contribution, and the interference between a splitting inside and a splitting outside, $\propto 2 \rmR \langle \Mc^\in_{\gamma \to q\bar q} \Mc^{\dag,\out}_{\gamma \to q \bar q} \rangle$ (second term), analogously referred to as an ``in-out'' contribution. Keeping the size of the medium fixed $L = \text{const.}$ and reducing the medium density $\hat q \to 0$ reveals a non-trivial cancellation between the two terms that leads to $F_\med \to 0$.

\section{Qualitative discussion of scales}
\label{sec:scales}

Presently, let us discuss the relevant scales that appear in the calculation. 
Considering \eqref{eq:MediumModification}, the emission process is characterized by a competition between the quantum mechanical formation process, that enforces $\tau \lesssim \tform$ for the ``in-in'' and $L-\ti \lesssim \tform$ for the ``in-out'' terms, respectively, as well as the suppression factors related to color decoherence and broadening. The condition on the splitting times is a consequence of avoiding strong oscillations of the trigonometric factors in \eqref{eq:MediumModification}.

The relevant scales for the ``in-in'' spectrum can be identified in the dipole and the factorizable piece of the quadrupole, i.e. the first term in \eqref{eq:FinalQuadrupole}. We will refer to them as the decoherence and broadening times, and they are given by
\beq
\tdecoh \sim \left(\frac{1}{\qhat \theta^2} \right)^{1/3} \,, \qquad \tbroad \sim \left(\frac{1}{\hat q \theta^2 L} \right)^{1/2}  \,.
\eeq
The non-factorizable part of the quadrupole constitutes a small correction to this qualitative estimate. The decoherence time governs the color decoherence of a dipole, and for $\tdecoh > L$, which implies that $\theta < \theta_c \sim (\hat q L^3)^{-1/2}$, the survival probability is close to one. This means that the medium does not resolve the dipole until it exits the medium. In particular, $\tform < \tdecoh$ implies that $\p^2 > \sqrt{\hat q z(1-z) E}$, which is related to the transverse momentum broadening accumulated during the formation time. The broadening time scale, on the other hand, is related to transverse momentum broadening along the medium length $L$. The condition $\tform < \tbr$ implies that $\p^2 > Q^2_s \sim \hat q L$. In the opposite case, the original opening angle of the dipole will be significantly changed by broadening, and the angle at which the particles emerge does not correspond to their initial opening angle. Note that $\tbr < \tdecoh$ for $\theta > \theta_c $, which implies that the broadening along the whole medium length is typically larger than during the formation time of (relatively) large-angle splittings. Hence, for emissions with $\theta < \theta_c$ one should not expect any medium modifications, i.e. $F_\text{med} = 0$. More importantly, the kinematical phase space for in-medium splittings that are vacuum-like, again implied by a vanishing $F_\text{med}$, also does exist for $\tform < \tbroad < \tdecoh$ at large angles $\theta > \theta_c$. We will compute the size of this phase space below up to leading-logarithmic precision.

It is important to keep in mind that the ``in-out'' term is not sensitive to the dynamics encoded in the quadrupole, see \eqn{eq:inout-semiclassic}. Instead, the spectrum is only sensitive to the decoherence time $\tdecoh$ in the dipole $S_{12}(L,t)$, see \eqref{eq:MediumModification}. At the same time, the phase limits the range of integration to $L-t \lesssim \tform$. Hence, it $\tform \ll L$ this term averages to 0. However, for $\tform \sim L \ll \tdecoh$ it cancels agains the ``in-in'' term. We will not discuss this contribution in further detail.

The discussion above holds for jets with $E > \omega_c \sim \hat q L^2$. For smaller energies, one finds stronger conditions on the angular phase space. Instead, one becomes sensitive to two dynamical critical angles given by $\tform \vert_{z=1} = \tdecoh$, leading to $\thetad \sim (\hat q/E^3)^{1/4}$, and $\tform \vert_{z=1} = \tbroad$, leading to $\thetabr \sim (\hat q L)^{1/2}/E$. Note that $E < \hat q L^2$ also implies that $\thetabr < \thetad$. Therefore, as long as $\theta_\text{br} > R$, or $E R > Q_s$, there exists a regime of hard, in-medium splittings. 

Hence, in order to avoid rapid oscillations or exponential suppression of the cross-section due to medium effects, the difference of emission times in the amplitude and its complex-conjugate of the in-in has to satisfy
\beq
\tau \lesssim \min[\tform,\tdecoh,\tbroad] \,,
\eeq
and the emission time for the in-in spectrum itself is of the same order $t \sim \tau$.
In other words, the actual emission time is governed by the smallest of the three physical time-scales of the problem.
Since $\tbroad < \tdecoh$ always, it turns out we can simply write $\tau \lesssim \min[\tform,\tbroad]$. Nevertheless, virtual emissions in the medium, that are critical to understand resummed observables in heavy-ion collisions \cite{Mehtar-Tani:2017web}, are not sensitive to final-state broadening and we will therefore continue to discuss full hierarchy of scales. At large formation times, $\tform \gg L$, we can neglect the factors in the integrands of Eqs.~\eqref{eq:inin-semiclassic} and \eqref{eq:inout-semiclassic}, to find that
\beq
\frac{\dd N^\text{in-in}}{\dd z \, \dd \theta}\big\vert_{\tform \gg L} \simeq \frac{\dd N^\vac}{\dd z \, \dd \theta} \times \left(\frac{L}{\tform} \right)^2 \qquad \text{and }\qquad \frac{\dd N^\text{in-out}}{\dd z \, \dd \theta}\big\vert_{\tform \gg L} \simeq - \frac{\dd N^\vac}{\dd z \, \dd \theta} \times \left(\frac{L}{\tform}  \right)^2 \,.
\eeq
We therefore expect that eventually $F_\med \approx 0$ at large formation times.

Based on the qualitative discussion of the medium spectrum presented above, we can draw a kinematical Lund diagram subject to the general constraint $p_\perp \equiv |\p| \geq Q_0$, where the cut-off scale $Q_0 \sim \Lambda_\text{QCD}$. In this work we choose to span the plane with the logarithmic variables $\{\ln\frac{1}{z}, \ln \frac{1}{\theta} \}$ and, for the sake of simplicity, we work in the double-logarithmic approximation (DLA) where we can neglect all corrections $\mathcal{O}(1-z)$, i.e. $p_\perp \simeq z E \theta$ etc., so that we only deal with straight lines in the Lund plane. This representation is well suited to detail the radiation pattern of soft and collinear emissions. The soft and collinear gluon emission off either a quark or gluon gives
\beq
\label{eq:DLA}
\frac{\dd \sigma^\text{\tiny DLA}}{\dd z \, \dd \theta} = \abar \frac{1}{z} \frac{1}{\theta} \,,
\eeq
where we defined $\abar \equiv  2\alpha_s C_R/\pi $ and where $C_R$ is the total color charge of the dipole. According to \eqref{eq:DLA}, at leading order the Lund plane is uniformly filled with density $\rho \sim \abar$ \cite{Andrews:2018jcm,Dreyer:2018nbf}.

At this stage it is worth pointing out that, although we have considered a photon splitting which does not contain any soft divergence, the factorization property of \eqn{eq:MediumSpectrum} and the general structure of the medium modification factor \eqn{eq:MediumModification} holds for an arbitrary splitting process. The generalization of our formulas to the splitting of a colored particle (quark or gluon), and the necessary replacements, will be further discussed in \autoref{sec:beyondsinglet}. At fixed coupling, the total phase space (PS) available for radiation off a jet with energy $E$ and a cone angle $R$ is therefore
\beq
(\text{PS})_\text{tot} =\frac{1}{\bar \alpha} \int_0^1 \dd z \int_0^R \dd \theta\, \frac{\dd \sigma^\text{\tiny DLA}}{\dd z \, \dd \theta} \Theta(zE \theta > Q_0) =  \frac{1}{2} \ln^2 \frac{E R}{Q_0} \,.
\eeq
In the presence of a medium, the four different competing time scales that we have identified are: (a) the kinematical formation time $\tform$, (b) the decoherence time $\tdecoh$, (c) the broadening time $\tbroad$ and (d) the medium length $\tend$. Note that in our present discussion all these timescales relate to the formation of the dipole and its further propagation through the medium. However, due to the fixed kinematics of the process, these timescales will also play a role in the further evolution of such a dipole, e.g. acting as a source for subsequent radiation.

\section{Mapping out the phase space for medium modifications}
\label{sec:LundDiag}

\begin{figure}[t!]
\centering
\includegraphics[width=0.49\textwidth]{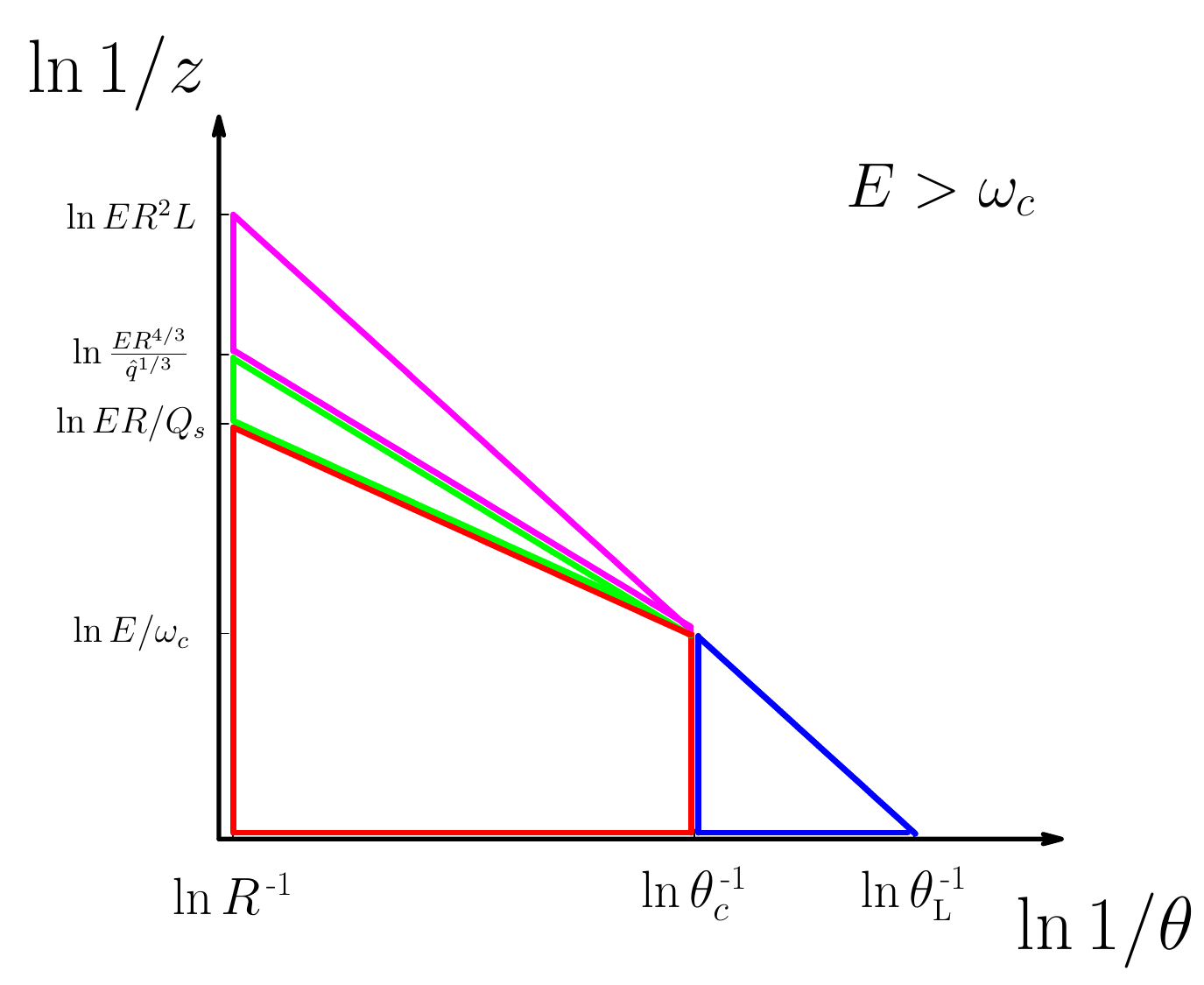}%
\includegraphics[width=0.49\textwidth]{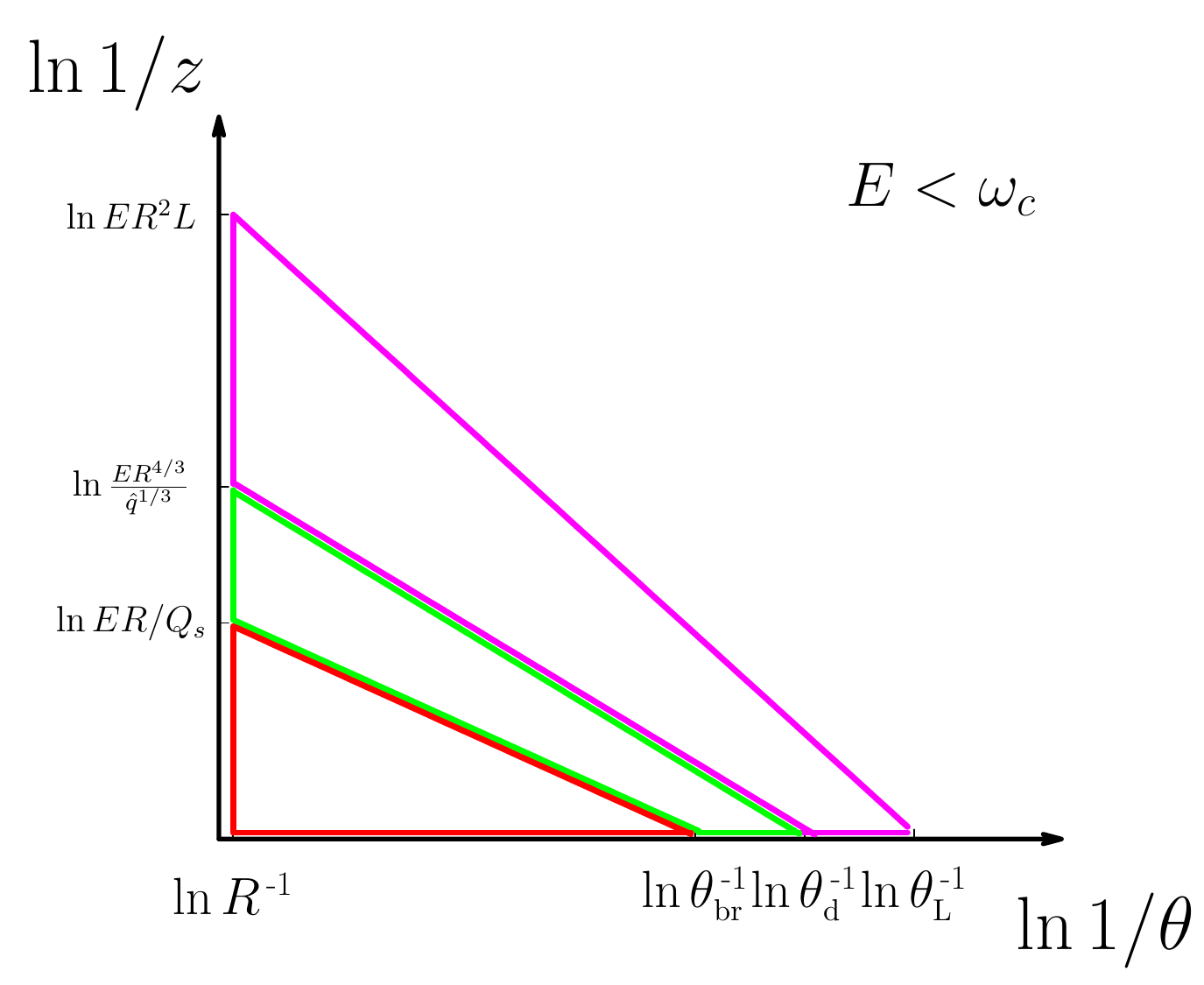}
\caption{Lund diagram for one vacuum splitting (soft \& collinear) where we detail the phase space regimes related to medium scales. All relevant definitions are given in \autoref{tab:scalesummary}.
}
\label{fig:LundDiagTheory} 
\end{figure}

We have sketched the Lund diagram for one in-medium splitting in \autoref{fig:LundDiagTheory} for two possible energy regimes: $E> \omega_c$ (left) and $E < \omega_c $ (right). The different ordering of the time-scales that were introduced above correspond to the marked areas on the graph and the lines are given by the following set of equations
\begin{align}
\label{eq:formation-time}
\ell &= \ln EL - 2 y \qquad (\tform = L) \,, \\
\ell &= \ln \frac{E}{\hat q ^{1/3}}- \frac{4}{3} y \qquad (\tform = \tdecoh)\,,\\
\ell &= \ln\frac{E}{Q_s} - y \qquad (\tform = \tbr) \,,
\end{align}
where $\ell \equiv \ln 1/z$ and $y \equiv \ln 1/\theta$. In the high-energy case $E> \omega_c$, we have also marked the critical angle $\theta_c$ with a vertical line. In this case, for $\theta < \theta_c$, $\tbroad > \tdecoh > L$ and so we have not extended the lines beyond their physical regime. 

\begin{table}[t]
\centering
\begin{tabular}{c | c }
Scale & Expression  \\
\hline
$Q_s$ & $(\hat q L)^{1/2} $ \\
$\omega_c$ & $\hat q L^2$ \\
$\theta_c$ & $ (\hat q L^3)^{-1/2}$ \\
$\thetad$ & $ (\hat q/E^3)^{1/4}$ \\
$\thetabr$ & $ (\hat q L)^{1/2}/E$ \\
$\thetaL $ & $ (E L)^{-1/2}$
\end{tabular}
\caption{Summary of the transverse momentum, energy and angular scales related to medium interactions.}
\label{tab:scalesummary}
\end{table}

Let us first consider the high-energy regime, i.e. $E > \omega_c$, see \autoref{fig:LundDiagTheory} (left). In order to avoid an interference between the edge of the medium and the non-perturbative scale $Q_0$, we also demand for now that $Q_0 < (RL)^{-1}$. For ease of reference, all the relevant scales are listed in \autoref{tab:scalesummary} with a short description.
Introducing the quantities $L_R \equiv \ln R/\theta_c$ and $L_E \equiv \ln E/\omega_c$, let us describe the various regions below:
\begin{description}
\item[\textbf{(A.1)} $\mathbf{ \tform < \tbroad <  \tdecoh < L}$ (red region):] Particles are created early in the medium, governed by the quantum mechanical formation time. This corresponds to vacuum-like emissions inside the medium, see also \cite{Mehtar-Tani:2017ypq,Caucal:2018dla}.
The phase space is given by
\beq
\label{eq:phase-space-1}
(\text{PS})_{1} =  L_R \left(L_E+ \frac{1}{2} L_R \right)  \,,
\eeq
and is single-logarithmic in the jet energy. The leading term arises for the case when $z E > \omega_c$. In fact, all other contributions are sub-leading in the jet energy as long as $\tdecoh < L$, see below, starting from the second term in \eqref{eq:phase-space-1}. Furthermore, $\tdecoh < L$ implies that the dipole will decohere in color in a finite distance inside the medium \cite{CasalderreySolana:2011rz,MehtarTani:2012cy}, which opens up for the possibility of incoherent energy loss due to secondary medium-induced radiation \cite{Mehtar-Tani:2017ypq}.

\item[\textbf{(A.2)} $\mathbf{\tbroad < \tform <  \tdecoh < L}$  (green region):] In this case, the timescale for broadening is shorter than the quantum mechanical formation time and we expect deviations from pure vacuum-like behavior. This region involves  relatively soft splittings, with $z E < \omega_c$ and $\p^2 < Q_s$, and the phase space reads
\beq
(\text{PS})_{2} =   \frac{1}{6} L^2_R \,,
\eeq
which is not enhanced by logs of the jet energy.

\item[\textbf{(A.3)} $\mathbf{\tbroad < \tdecoh < \tform < L }$ (magenta region):] The formation of the dipole in this region is strongly suppressed by the presence of the dipole governing the decoherence of the pair before formation. In particular, this time ordering implies that $\omega/L < \p^2 < (\qhat zE )^{\onehalf}$. The phase space is 
\beq
(\text{PS})_{3}= \frac{1}{3} L^2_R \,,
\eeq
and is also not enhanced by logs of the jet energy.

\item[\textbf{(A.4)} $ \mathbf{\tform < L < \tdecoh < \tbroad}$  (blue region):] In this case, the splitting takes place inside the medium, but the created partons remain coherent. This happens if the splitting angle is sufficiently small, $\theta < \theta_c$ \cite{MehtarTani:2010ma,MehtarTani:2011tz}. This  implies further that splittings in this region should follow a vacuum emission pattern. The phase space is given by
\beq
(\text{PS})_{4} = \frac{1}{4} L_E^2 \,,
\eeq
and is double-logarithmic in the jet energy. In this case the jet is quenched (coherently) due to the presence of a total color charge \cite{CasalderreySolana:2012ef,Mehtar-Tani:2017ypq}. It is worth pointing out that this regime does not exist in the low-energy regime, $E < \omega_c$. Due to the restriction on the energy, the characteristic decoherence and broadening times are always shorter than the medium length and, in effect, all radiation inside the medium, i.e. $\tform < L$, occur at angles $\theta > \theta_c$, since $(\omega_c \theta^2)^{-1} < (zE \theta^2)^{-1} < L$.

\item[\textbf{(B)} $\mathbf{\tform > L }$ (beyond the $\tform = L$ line):] Splitting takes place outside of the medium, and no medium modification is expected.
\end{description}
To summarize, we have identified two regions of vacuum-like emissions inside the medium, $\tform < L$, namely regions {\bf A.1} and {\bf A.4}. However, the fate of the dipole after splitting is expected to be very different. We therefore denote region {\bf A.1} as incoherent radiation and region {\bf A.4} as coherent radiation.

The total phase space for emissions inside the medium is the sum of the four contributions,
\beq
(\text{PS})_{\tform < L} \,=\, \sum_{i=1}^4(\text{PS})_i \,=\,  \frac{1}{4} \ln^2 E R^2 L \,.
\eeq
Note therefore that there is a relatively large probability of a splitting happening inside the medium, i.e. $(\text{PS})_{\tform <L}/(\text{PS})_\text{tot} \sim 1/2$ asymptotically when $E \to \infty$. In Monte Carlo simulations, the ratio is slowly varying and lies close to $\sim$40$-$45\% \cite{CasalderreySolana:2011gx}. 

Finally, let us briefly consider the low-energy regime, i.e. $E < \omega_c$. In this case, the logarithmic contributions are automatically restricted and the leading-logarithmic approximations should receive significant corrections. However, it is interesting to note that the region $\tform < \tbroad < \tdecoh$, corresponding to region {\bf A.1} in the high-energy regime, scales like 
\beq
(\text{PS})_{\tform < \tbroad, E< \omega_c} = \frac{1}{2} \ln^2 \frac{E R}{Q_s} \,.
\eeq
As pointed out before, the regime of in-medium hard splittings closes whenever $E R \sim Q_0$. We will currently not discuss in further detail the remaining phase space regimes, although their impact can be systematically worked out following the steps above.

\section{Numerical results}
\label{sec:numerics}

\begin{figure}[t!]
\centering
\includegraphics[width=0.48\textwidth]{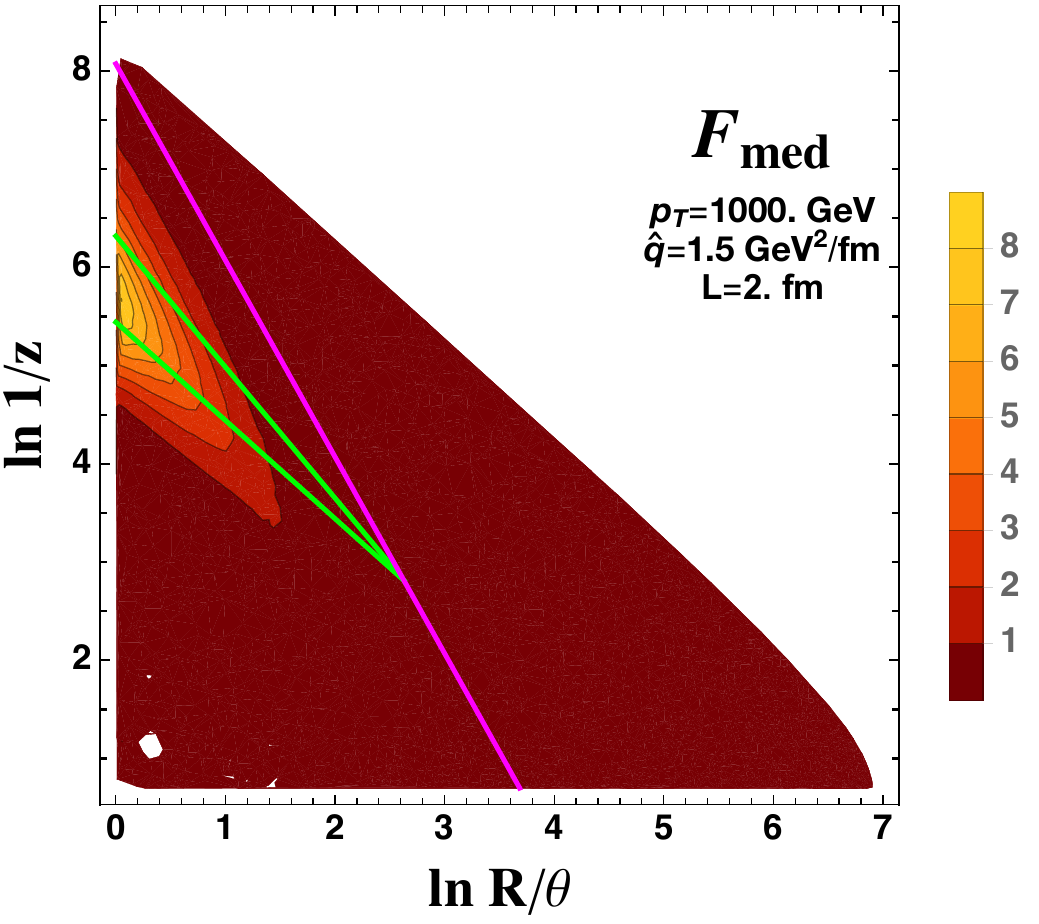}
\includegraphics[width=0.48\textwidth]{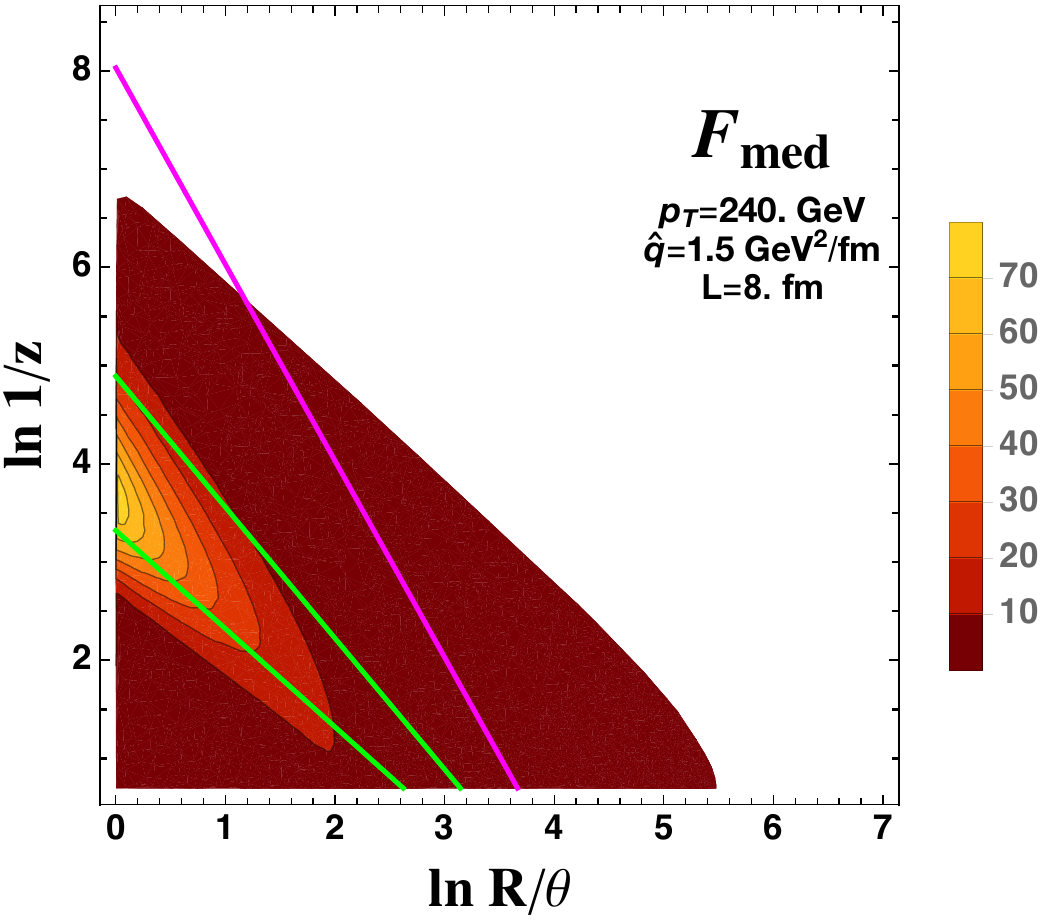}
\caption{Numerical evaluation of the medium modification factor $F_\med$ for in the high-energy ($ E> \omega_{c} $, left) and low-energy regime ($E < \omega_c$, right) . Notice that the scale of the color coding on the right is rescaled by a factor 10 compared to the left. The shaded area corresponds to the available phase space given the constraint $k_\perp > Q_0$.}
\label{fig:LundDiagNew}
\end{figure}

The main result of this work is to demonstrate the factorization property of the medium spectrum given in \eqref{eq:MediumSpectrum}. We have chosen $Q_0 = 0.2$ GeV and $\hat q = 1.5$ GeV$^2$/fm as reference values, for the high-energy regime we have used $E = 1000 $ GeV and $L = 2$ fm and for the low-energy regime we have chosen $E = 240$ GeV and $L = 8$ fm.\footnote{The reason behind this choice is to plot the same in-medium phase space $\tform < L$ so that the characteristic angle $\theta_c$ in the high-energy regime is located approximately at the same absolute angle as $\thetabr$ in the low-energy regime.} Our final results are presented in \autoref{fig:LundDiagNew}, where the quantity $F_\med$ is defined in \eqn{eq:MediumModification}. We have plotted the result of evaluating the medium-modification function $F_\med$ in the kinematical Lund diagram defined in \autoref{fig:LundDiagTheory} for both high- and low-energy regimes in \autoref{fig:LundDiagNew} (left) and \autoref{fig:LundDiagNew} (right) respectively, where the lines in the two figures are equivalent.\footnote{We have confirmed that the borders do not shift by any significant amount if we were to include numerical factors into the various scales used throughout, adding to the robustness of the DLA analysis.} The full shaded area corresponds to the available phase space given the constraint $k_\perp > Q_0$, in such a way that the three curves represent (from top to bottom, cf. \autoref{fig:LundDiagTheory}): a) $\tform = L$, b) $\tform = \tdecoh$ and c) $\tform = \tbroad$.

It is instructive to examine how the medium modification function behaves for different limits. Commencing our discussion with the high-energy regime, see \autoref{fig:LundDiagNew} (left), the regimes where we expect vacuum-like emissions to occur, i.e. at $\tform < \tbroad < L$ with $\tdecoh > L$ ($\theta < \theta_c$) and $\tform > L$, the medium modifications are negligible. Indeed, we observe that the onset of modifications follow the line $\tform = \tdecoh$ and the main modifications are contained to the regime $\tbroad < \tform < \tdecoh$, as expected from the discussion in \autoref{sec:scales}. This behavior is more striking, the smaller the coherence angle compared to the cone size. This is of little surprise given that $\tbroad $ is related to transverse momentum broadening along the medium length $ L $, hence making sense for larger media.

In the low-energy regime, see \autoref{fig:LundDiagNew} (right), the same physical picture holds to a large extent. However, the medium modifications are much larger and we also observe a ``leakage'' into the regime of short formation times. In this case, the scaling behavior we have postulated can only be thought to hold in a parametric sense, and care has to be taken with the assumptions regarding the importance of transverse momentum broadening in order to make any quantitative statements. In \autoref{fig:LundDiagNew} (right), the parameters are such that $Q_0 > (RL)^{-1}$. Apart from serving to prove the expected scaling in the low-energy regime, this parameter choice illustrates that part of the jet, i.e. large-angle and soft emissions, happen to reach the non-perturbative scale while still being ``inside'' the medium, i.e. their formation times being smaller than $L$. This constitutes a new category of in-medium modifications that goes beyond the scope of our investigation.

\section{Beyond the singlet case}
\label{sec:beyondsinglet}

The generalization of the splitting process to arbitrary color representation is straightforward, and does not modify the general structure of \eqn{eq:MediumSpectrum}. One simply has to replace the coupling constant $\alpha_\text{em} \to \alpha_s$ and the Altarelli-Parisi splitting function for the relevant one, e.g. $P_{q\gamma}(z) \to P_{ij}(z)$ in the vacuum spectrum \eqref{eq:VacuumSpectrum}. In the medium modification factor $F_\med$ one has to account for the more involved color algebra. Taken as a concrete example the splitting process $q \to q+g$, we would replace the expressions for the dipole and quadrupole by,
\begin{align}
S_{12} (\tf,\ti) &\to \frac{1}{N_c^2-1}\left\langle \rmtr \left(V_2^\dag(\ti,\tf) V_1(\tf,\ti)\right) \rmtr\left(V_0^\dag(\ti,\tf) V_2(\tf,\ti)\right) - \frac{1}{N_c}\rmtr\left(V_0^\dag(\ti,\tf) V_1(\tf,\ti)\right) \right\rangle \,, \\
Q &\to \frac{1}{N_c^2-1}\left\langle \rmtr \left(V_{\bar 1}^\dag(\tf,\tend) V_1(\tend,\tf) V_2^\dag(\tf,\tend) V_{\bar 2}(\tend,\tf)\right) \rmtr\left(V_{\bar 2}^\dag(\tf,\tend) V_2(\tend,\tf)\right) \right. \nonumber \\
&\qquad \quad \left. -\frac{1}{N_c}\rmtr\left(V_{\bar 1}^\dag(\tf,\tend) V_1(\tend,\tf)\right) \right\rangle \,.
\end{align}
In the large $N_c$ limit, the second term in both expressions above can be neglected, in which case we are left with the replacement
\begin{align}
S_{12} (\tf,\ti) &\to S_{12}(\tf,\ti) S_{20}(\tf,\ti) \,, \\
Q &\to Q(\tend,\tf) S_{2\bar 2}(\tend,\tf)\,.
\end{align}
The appearance of these new dipole structures, $ S_{20}(\tf,\ti) $ and $ S_{2\bar 2}(\tend,\tf) $, does not render it impossible to transfer the qualitative insight regarding color-singlet dipole splittings to color-charged ones. In the harmonic oscillator approximation, we find that
\begin{align}
S_{20} (\tf,\ti) &=\rme^{-\frac{1}{12}\hat{q}(1-z)^{2}\theta^{2} \tau^{3}}, \\
S_{2 \bar{2}} (L,\tf,\ti) &= \rme^{-\frac{1}{4}\hat{q}z^{2}\theta^{2}(L-t') \tau^{2}},
\end{align}
where we have defined $ \r_{2 0}(s) = \n_1 (s- \ti) $ and $ \r_{2 \bar{2}}(s) = \n_2 (t'- \ti) $.

Focussing for the moment on the dynamics during formation, we note that the combination $S_{12}S_{20}$ only depends on the jet quenching parameter through the combination $\hat q_\text{eff} = \hat q (1 + (1-z)^2) \approx ((1-z) N_c + z^2 C_F ) \hat{ \bar q}$ in the large-$N_c$ limit (recall that, in our calculation so far, $\hat q = C_F \hat{\bar q}$). This effective $\hat q_\text{eff}$ was indeed previously identified for medium-induced quark-gluon splitting, see e.g. \cite{Apolinario:2014csa}.

Similar conclusions can be drawn regarding the correction to the quadrupole. Therefore, although our general discussion was based on the calculation of a color-singlet splitting, we see that it can be straightforwardly generalized to splittings involving a non-zero total color charge by carefully considering the color dependence of $\hat q_\text{eff}$, as well as the expected replacements in the vacuum spectrum. This validates our discussion of soft and collinear radiation in terms of the Lund plane introduced in \autoref{sec:LundDiag}.

\section{Conclusions and outlook}
\label{sec:conclusions}

We have studied the production of hard radiation in the presence of a quark-gluon plasma, and have found two regimes of vacuum-like emissions inside the medium. By vacuum-like, we simply mean that the in-medium splitting function is equal to the one in the vacuum, or $F_\med \approx 0$. These cases include a) the regime of short formation time, in particular $\tform < \tbr < L$ (corresponding to the region {\bf A.1} in \autoref{fig:LundDiagTheory}), and at small angles, concretely $\theta < \theta_c$ (corresponding to region {\bf A.4} in \autoref{fig:LundDiagTheory}). The fate of these two types of emissions is however different since only in the former regime do the splitting products lose their color coherence at a finite distance inside the medium. 
Due to this rapid decoherence, the splitting products should therefore become subject to independent energy loss processes. In contrast, emissions in region {\bf A.4} are  color coherent when they exit the medium and should therefore lose energy as a whole.

We have also identified the border at which long-distance medium effects start to play a role in the splitting process. In particular at $\tform \gtrsim \tbr$, where the relative transverse momentum $\p^2 < Q_s^2$, the opening angle of the jet could vary significantly due to transverse momentum broadening. These features are also recovered in our numerical calculations in \autoref{sec:numerics}. Hence, this study confirms the notion of purely vacuum-like emissions that are emitted inside the medium. 

The spectrum of these excitations follow from a ``semi-classical'' picture, where the original splitting takes place immediately after the hard process. Our analysis of time-scales further corroborates the validity of our assumptions for the regions of vacuum-like emissions. A further, quantitative study of the regions where $\tform$ is similar to $\tbroad$ and $\tdecoh$ demands that we include the possibility of transverse momentum broadening, i.e. relax the assumption of straight-line trajectories, for further details see \cite{Blaizot:2012fh,Apolinario:2014csa}.

Computing higher-order splitting processes, and their virtual corrections, goes beyond the scope of this paper, and will be pursued in the future. We already anticipate that our analysis points to mismatch between real and virtual terms, since the latter do not involve the long-time component of the processes encoded, in our case, in the quadrupole. Such a mismatch, albeit due to energy loss processes, was already shown to entail novel resummation schemes to account for medium modifications on multi-parton probes, such as jets \cite{Mehtar-Tani:2017web}.
 
\section*{Acknowledgements}
KT is supported by a Starting Grant from Trond Mohn Foundation (BFS2018REK01) and the University of Bergen.
VV is supported by the Spanish FPU Fellowship FPU16/02236;
CAS and VV are supported by Ministerio de Ciencia e Innovaci\'on of Spain under project FPA2017-83814-P and Unidad de Excelencia Mar\'\i a de Maetzu under project MDM-2016-0692, by Xunta de Galicia (Conseller\'\i a de Educaci\'on) and FEDER. JGM is supported by Funda\c  c\~ao para a Ci\^encia e a Tecnologia (Portugal) under project CERN/FIS-PAR/0022/2017, and he gratefully acknowledges the hospitality of the CERN theory group.

\appendix
\section{Beyond the classical picture}
\label{sec:beyondclassical}

Applying the limit $E \to \infty$ and using the previously derived dressed propagator in  \eqn{eq:propagator-nobroad}, we obtain the amplitude,
\begin{align}
\mathcal{M}^{\rm in}_{\gamma \rightarrow q \bar{q}} &= \frac{e}{E} \gamma^{\gamma\to q\bar q}_{\lambda,s,s'} (z) \,\int_0^\tend \dd \ti \, \rme^{-i(\tend- t)/\tform} \nn
&\times \Big[\p + i[(1-z) \bdel_{x_1} - z \bdel_{x_2}] \Big] \cdot \beps_\lambda [V_{1}(t_{L}, \ti) V^{\dagger}_{2}(t_{L},\ti)]_{ij} \big\vert_{\x_1=\x_2 = \n_0 t} \,,
\end{align}
up to factors that cancel out in the cross section, and where $\n_0 = (\p_1+\p_2)/E$. Note that, in this case, the trajectories of the dipole constituents are described by $\r_i \equiv \x_i+ (s-t)\n_i$, while in \autoref{sec:semi-classical} we assumed that $\x_i =0$.

The ``in-in'' and ``in-out'' emission spectra then become
\begin{align}
\label{eq:inin-full}
\frac{\dd N^\text{in-in}}{\dd z \, \dd \p^2} &= \frac{\dd N^\text{vac}}{\dd z \, \dd \p^2}\; 2 \rmR \int_0^\tend \frac{\dd \ti}{\tform} \int_{\ti}^\tend \frac{\dd \tf}{\tform} \, \rme^{-i \frac{\tf - \ti}{\tform}}  \,\mathcal{\hat V}_1 \big[Q (\tend, \tf) S_{12}(\tf,\ti) \big]_{\substack{\x_2=\x_1 = \n_0 \ti \\ \bar \x_2 = \bar \x_1 = \n_0 \tf}} \,,
\end{align}
and 
\beq
\label{eq:inout-full}
\frac{\dd N^\text{in-out}}{\dd z \, \dd \p^2} = \frac{\dd N^\text{vac}}{\dd z \, \dd \p^2}\; 2 \text{Im} \int_0^\tend \frac{\dd \ti}{\tform} \, \rme^{-i \frac{\tend-\ti}{\tform}} \mathcal{\hat V}_2 \big[ S_{12}(L,\ti) \big]_{\x_2=\x_1 = \n_0 \ti} \,,
\eeq
where $\tform = 2 z (1-z) E/\p^2$ and we have introduced the operators
\begin{align}
\label{eq:vertex-inin}
\mathcal{\hat V}_1 &= \frac{1}{\p^2}\Big(\p + i[(1-z) \bdel_{x_1} - z \bdel_{x_2} ] \Big) \cdot \Big(\p - i[(1-z) \bdel_{\bar x_1} - z \bdel_{\bar x_2} ] \Big) \,, \\
\label{eq:vertex-inout}
\mathcal{\hat V}_2 &=  \frac{\p}{\p^2} \cdot \big(\p + i[(1-z) \bdel_{x_1} - z \bdel_{x_2}] \big) \,.
\end{align}
The dipole term, comprising the additional shift of the initial positions of the Wilson lines, reads
\beq
S_{12}(t_1,t_0) = \exp \left\{-\frac{1}{4}\hat q \Delta t \left[ \left(\x_{12}+ \frac{1}{2}\Delta t \n_{12} \right)^2 + \frac{1}{12} \Delta t^2 \n_{12}^2\right] \right\} \,,
\eeq
for generic time intervals, where $\Delta t = t_1 - t_0$ and $\x_{12}\equiv \x_1 - \x_2$, while the missing terms in the quadrupole \eqref{eq:quadrupole} read, explicitly $S_{I\bar I}(t_1,t_0) = \exp [-\frac{1}{4} \hat q \Delta t \, (\x_{I\bar I} + \n_I \tau)^2]$, using the definition in \eqref{eq:hoa}.

Because of the constraints on the initial transverse position in the amplitude and the complex conjugate, leading respectively to $\x_1 = \x_2$ and $\bar \x_1 = \bar \x_2$, the resulting spectra will be similar to the terms derived to obtain \eqn{eq:MediumSpectrum}, with the definition in \eqref{eq:MediumModification}, except for a unique pre-factor appearing under the integrals of the ``in-in'' and the ``in-out'' terms that arises from the more involved vertices in \eqref{eq:inin-full} and \eqref{eq:inout-full}.

We have analyzed these terms in detail for the factorizable piece of the ``in-in'' term and for the ``in-out'' term. In particular, for the ``in-in'' term the correction factor appearing under the integral reads
\beq
\label{eq:corrections-1}
1 - i\frac{\hat q \tau^2}{4 z(1-z)E} -i \frac{\hat q \tau_L \tau \xi}{z(1-z) E} -\left(\frac{\hat q \tau_L \tau \xi}{2z(1-z) E} \right)^2 \left(1 + \frac{\tau}{2 \xi \tau_L} \right) + \frac{\hat q \tau_L \xi}{(z(1-z)E \theta)^2} \, ,
\eeq
where we defined $\tau_L\equiv L-\tf$ and $\xi \equiv (1-z)^2 + z^2$ to shorten the expression. For the in-out term the correction factor is $1- i \hat q (L-\ti)^2/[4 z(1-z)E]$, which closely resembles the two first terms in \eqref{eq:corrections-1} with $\tau$ substituted by $L-\ti$. Neglecting all finite-$z$ corrections and assuming short times, $L \gg \ti, \tf$, these terms scale as
\beq
\label{eq:corrections-scaling}
1-i \frac{\tform}{\tdecoh}\left(\frac{\tau}{\tdecoh} \right)^2 -i \frac{\tform}{\tbroad}\frac{\tau}{\tbroad} - \left(\frac{\tform}{\tbroad} \right)^2 \left(\frac{\tau}{\tbroad} \right)^2 + \left(\frac{\tform}{\tbroad} \right)^2 \,.
\eeq
This clearly demonstrates that the corrections to the vertex start to play a role whenever the (kinematical) formation time ceases to constitute the shortest time-scale to which compare the difference of emission times $\tau$. In particular, this starts to happen then $\tau \lesssim \tbr < \tform$.

\bibliographystyle{elsarticle-num}
\bibliography{jetquenching}

\end{document}